\documentclass[twocolumn,aps,prl,superscriptaddress]{revtex4-2}
\usepackage{color}
\usepackage{mathrsfs}
\usepackage{bm}
\usepackage{amsmath}
\usepackage{amssymb}
\usepackage{graphicx}
\usepackage{hyperref}
\usepackage{color}
\usepackage{siunitx}

\makeatletter
\hypersetup{hypertex=true, colorlinks=true, linkcolor=blue, anchorcolor=blue,citecolor=blue}

\usepackage{dcolumn}
\usepackage{bm}
\usepackage{mathrsfs}
\usepackage{amsfonts}
\usepackage{xcolor}
\usepackage{epstopdf}

\begin{document}
\title{Generalized Topology in Lattice Models without Chiral Symmetry}
\author{Qing Wang}
\affiliation{Anhui Province Key Laboratory of Low-Energy Quantum Materials and
Devices, High Magnetic Field Laboratory, HFIPS, Chinese Academy
of Sciences, Hefei, Anhui 230031, China}
\affiliation{Science Island Branch of Graduate School, University of Science and
	Technology of China, Hefei, Anhui 230026, China}
\author{Ning Hao}
\email{haon@hmfl.ac.cn}
\affiliation{Anhui Province Key Laboratory of Low-Energy Quantum Materials and
Devices, High Magnetic Field Laboratory, HFIPS, Chinese Academy
of Sciences, Hefei, Anhui 230031, China}

\begin{abstract}
The Su-Schrieffer-Heeger (SSH) model is a fundamental lattice model used to study topological physics. Here, we propose a new versatile one-dimensional (1D) lattice model that extends beyond the SSH model. Our 1D model breaks chiral symmetry and has generalized topology characterized by a projected winding number $W_{1D,P}=1$. When this model is extended to 2D, it can generate a second-order topological insulator (SOTI) phase. The generalized topology of the SOTI phase is protected by a pair of opposite winding numbers $W_{2D,P}^{\pm}=\pm1$, which count the opposite phase windings of a projected vortex and antivortex pair defined in the manifold of the entire parameter space. Thus, the topology of our models is robust and the end (corner) modes are independent of the selection of unit cells and boundary configurations. More significantly, we demonstrate that the model is very general and can be inherently realized in many categories of crystalline materials such as BaHCl.
\end{abstract}
\maketitle

\textit{Introduction}.-Topological insulators have been extensively researched in many fields of physics due to their distinctive properties \cite{PhysRevLett.95.226801,BHZ,RevModPhys1,RevModPhys2,RevModPhys3}. Various effective fundamental models have been established, such as the Haldane model \cite{PhysRevLett.61.2015}, the Qi-Wu-Zhang model \cite{QWZ}, the Bernevig-Hughes-Zhang model \cite{BHZ} and the Kane-Mele model \cite{PhysRevLett.95.146802} for simple lattices. In recent years, there has been a further expansion into higher-order topological phases \cite{BBH,PhysRevB.96.245115,sciadv.aat0346,PhysRevLett.119.246402,PhysRevB.98.081110,PhysRevB.97.205136,PhysRevB.97.205135,PhysRevB.99.041301,PhysRevLett.123.256402,PhysRevLett.124.166804,PhysRevLett.126.066401,PhysRevB.107.235406}. Many theoretical proposals based on these concepts have been implemented across various experimental platforms in different fields \cite{PhysRevLett.124.206601,PhysRevB.99.020304,NaturePhysics,Nature,mittal2019photonic,serra2018observation}.  For a wider range of complex lattices, the well-known SSH model \cite{SSH,SSH2} and the Rice-Mele model \cite{RM} provide the foundation to build effective models for high-order topological phases. For instance, the Benalcazar-Bernevig-Hughes (BBH) model \cite{BBH} and numerous 2D-SSH models \cite{PhysRevLett.118.076803,PhysRevB.100.235137,PhysRevB.107.045118,PhysRevB.106.245109,10.3389/fphy.2022.861242,PhysRevB.108.245140,PhysRevB.109.195154} have been developed to describe the SOTI phases. 

The nontrivial topology of these models requires that intercell hopping is larger than intracell hopping. Consequently, the emergence of topological end states or corner states depends on the choice of the unit cell and specific boundary configurations. In practice, the required pseudo-dimer structures \cite{BBH,PhysRevLett.118.076803,PhysRevB.100.235137,PhysRevB.107.045118,PhysRevB.106.245109,10.3389/fphy.2022.861242,PhysRevB.108.245140,PhysRevB.109.195154} are seldom energetically favorable, and the specific boundary configurations \cite{BBH,PhysRevLett.118.076803,PhysRevB.100.235137,PhysRevB.107.045118,PhysRevB.106.245109,10.3389/fphy.2022.861242,PhysRevB.108.245140,PhysRevB.109.195154} are also not the natural cleavage boundaries in most materials. These constrains limit the search for materials exhibiting the high-order topological phases based on the aforementioned models. Furthermore, most of the predicted higher-order topological materials do not possess the pseudo-dimer structure described in the above models. As a result, their topology can only be characterized through the method of analyzing the occupied state parities at specific high-symmetry points \cite{sciadv.aat0346,PhysRevB.99.245151}. Therefore, there is a need to construct more universally applicable lattice models, independent of boundary configurations and easy to implement, to guide the search for high-order topological materials  and to intuitively understand their topological physics.

 \begin{figure}[pb]
	\begin{center}
		\includegraphics[width=1.0\columnwidth]{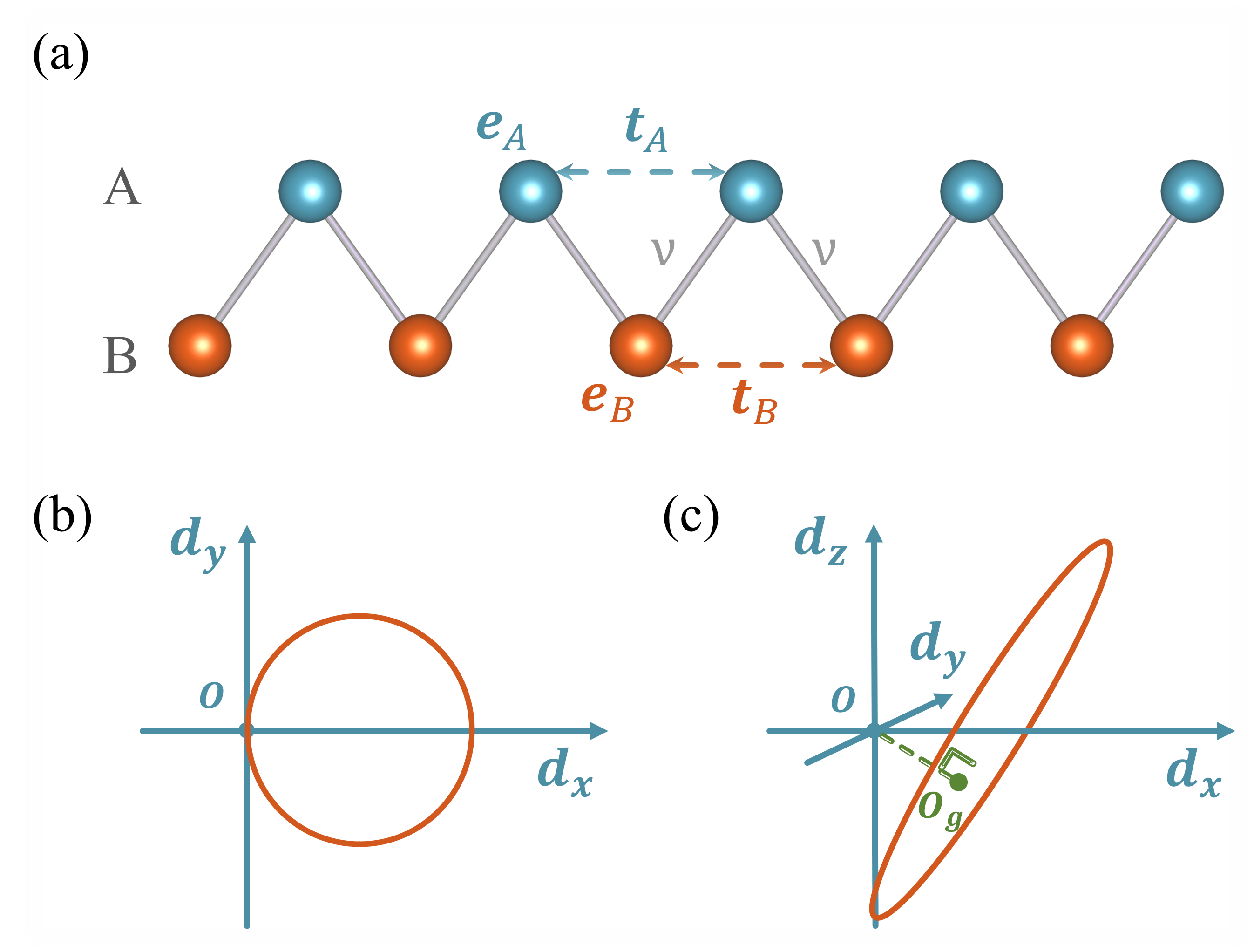}
	\end{center}
	\caption{(a) Schematic of a 1D atomic chain with on-site energy $e_{A}$($e_{B}$) of sublattice A(B), nearest-neighbor hopping $\nu$, and next-nearest-neighbor hopping  $t_{A}$($t_{B}$) of sublattice A(B). (b-c) The trajectory of the endpoint of $\boldsymbol{d}(k)$ vector in $d_{x}(k)-d_{y}(k)$ plane (b) and three-dimensional space (c). When parameter $k$ evolves from 0 to $2\pi$. The projection origin $O_{g}$ lies inside the closed curve, the system is non-trivial in (c).}%
	\label{fig-1}%
\end{figure}

In this work, we construct a universal 1D two-sublattice atom chain model with formally equal intracell and intercell hoppings, and introduce the concept of generalized topology through the method of projected winding numbers to characterize the topology of the model. Using this 1D model as a buliding block, we construct a model in the 2D case. We find that the 2D model can support a SOTI phase characterized by a pair of opposite winding numbers $W_{2D,P}^{\pm}=\pm1$, which count the opposite phase windings of a projected vortex and antivortex pair defined in the manifold of the entire parameter space. The topological phase transitions are characterized by the creation or annihilation of the projected vortex-antivortex pair in the parameter space during the evolution of the system. Remarkably, the corner modes from the SOTI phase are independent of the specific boundary configurations. More meaningfully, we show that the models can be inherently realized by many categories of crystalline materials such as BaHCl.   

\textit{1D model and topology}.-We consider a 1D atomic chain with A and B sublattices (Fig. \ref{fig-1}(a)). Up to the next-nearest-neighbor hopping, its Hamiltonian in the momentum-space can be expressed as:
\begin{align}
	H(k) =\begin{pmatrix}
		e_{A}+2 t_{A}\cos(k) & \nu+e^{-ik}\nu \\
		\nu+e^{ik}\nu & e_{B}+2 t_{B}\cos(k)
	\end{pmatrix}.
\end{align}
Here, $e_{A}$($e_{B}$) represents the on-site energy of sublattice A(B), $\nu$ denotes nearest-neighbor hopping amplitude, and $t_{A}$($t_{B}$) represents the next-nearest-neighbor hopping of sublattice A(B). It can be further expressed in a compact form, $H(k) =\boldsymbol{d}(k)\cdot\boldsymbol{\sigma}$ with a four-component $\boldsymbol{d}$ vector and four Pauli matrices spanned in sublattice space. The energy eigenvalue $E_{\pm}(k)=d_{0}(k)\pm\sqrt{d_{x}^2(k)+d_{y}^2(k)+d_{z}^2(k)}$. Note that the $d_{0}(k)\sigma_{0}$ term has no impact on the value of $\bigtriangleup E(k)=E_{+}(k)-E_{-}(k)$ and the relevant wavefunctions \cite{PhysRevB.99.035146,PhysRevLett.127.147401}, which contains the topological information. One can neglect this term and the effective Hamiltonian in the momentum-space can be expressed as:
\begin{align}
	H(k)	=\begin{pmatrix}
e+2t\cos(k) & \nu+e^{-ik}\nu \\
\nu+e^{ik}\nu & -e-2t\cos(k)
	\end{pmatrix}.
\end{align}
Here, $e=(e_{A}-e_{B})/2$,  $t=(t_{A}-t_{B})/2$. In the absence of $d_{z}(k)$ term, the trajectory of the endpoint of $\boldsymbol{d}(k)$ vector forms a circle passing through the origin in $d_{x}(k)-d_{y}(k)$ plane when the parameter $k$ evolves from 0 to $2\pi$, as shown in Fig. \ref{fig-1}(b). This corresponds to the critical point in the SSH model \cite{SSH,SSH2}, where winding number is ill-defined. In the presence of $d_{z}(k)$, the trajectory of the endpoint of $\boldsymbol{d}(k)$ vector extends to a closed curve in three-dimensional space ($d_{x}(k), d_{y}(k), d_{z}(k)$), as shown in Fig. \ref{fig-1}(c). It's clear that the breaking of chiral symmetry due to nonzero $d_{z}(k)$ term indicates that the plane containing the closed curve cannot pass through the origin $O$, and the standard winding number cannot be defined. 

To capture the hidden topological properties in the presence of chiral-symmetry-breaking $d_{z}$ term, we introduce a projected winding number,
 \begin{align}
	W_{1D,P}=\frac{1}{2 \pi} \int_{0}^{2\pi}\left({\frac{\mathbf{g}(k)}{|\mathbf{g}|}} \times \frac{d}{d k} {\frac{\mathbf{g}(k)}{|\mathbf{g}|}}\right) d k.
\end{align}
Here, $\mathbf{g}(k) = \hat{P} \mathbf{d}(k)$ represents the projected 2D $g_{x}(k)-g_{y}(k)$ plane, where the projection operator $\hat{P}$ is defined as follows:
\begin{align}
	\hat{P} =\begin{pmatrix}
		\frac{\nu}{\sqrt{4 t^2+\nu^2}} & 0 & \frac{2 t}{\sqrt{4 t^2+\nu^2}} \\
		0 & 1 & 0 \\
	\end{pmatrix}.
\end{align}
\begin{figure}[pt]
	\begin{center}
		\includegraphics[width=1.0\columnwidth]{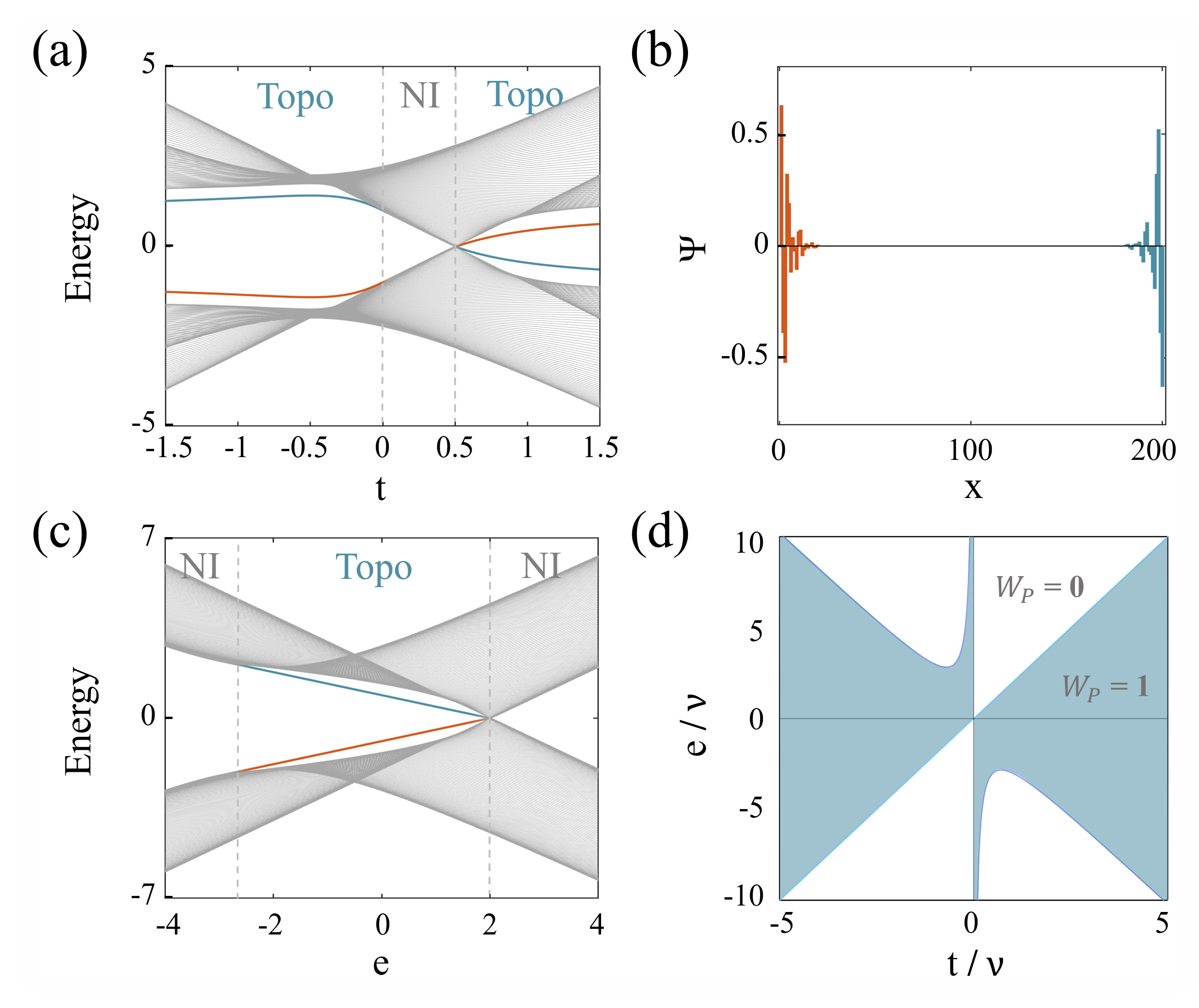}
	\end{center}
	\caption{(a) The energy spectrum of a 1D atomic chain with a length of 100 unit cells evolves with changes in next-nearest-neighbor hopping $t$. The system is trivial for $0<t<0.5$, and nontrivial for $t<0$ or $t>0.5$. (b) The spatial distribution of end states corresponding to (a). (c) The energy spectrum evolves with changes in staggered on-site energy $e$. (d) Phase diagram illustrating the system's evolution with parameters. The projected winding number $W_{P}=1$ represents a non-trivial topological phase. }%
	\label{fig-2}%
\end{figure}

\begin{figure*}[tph]
	\begin{center}
		\includegraphics[width=2\columnwidth]{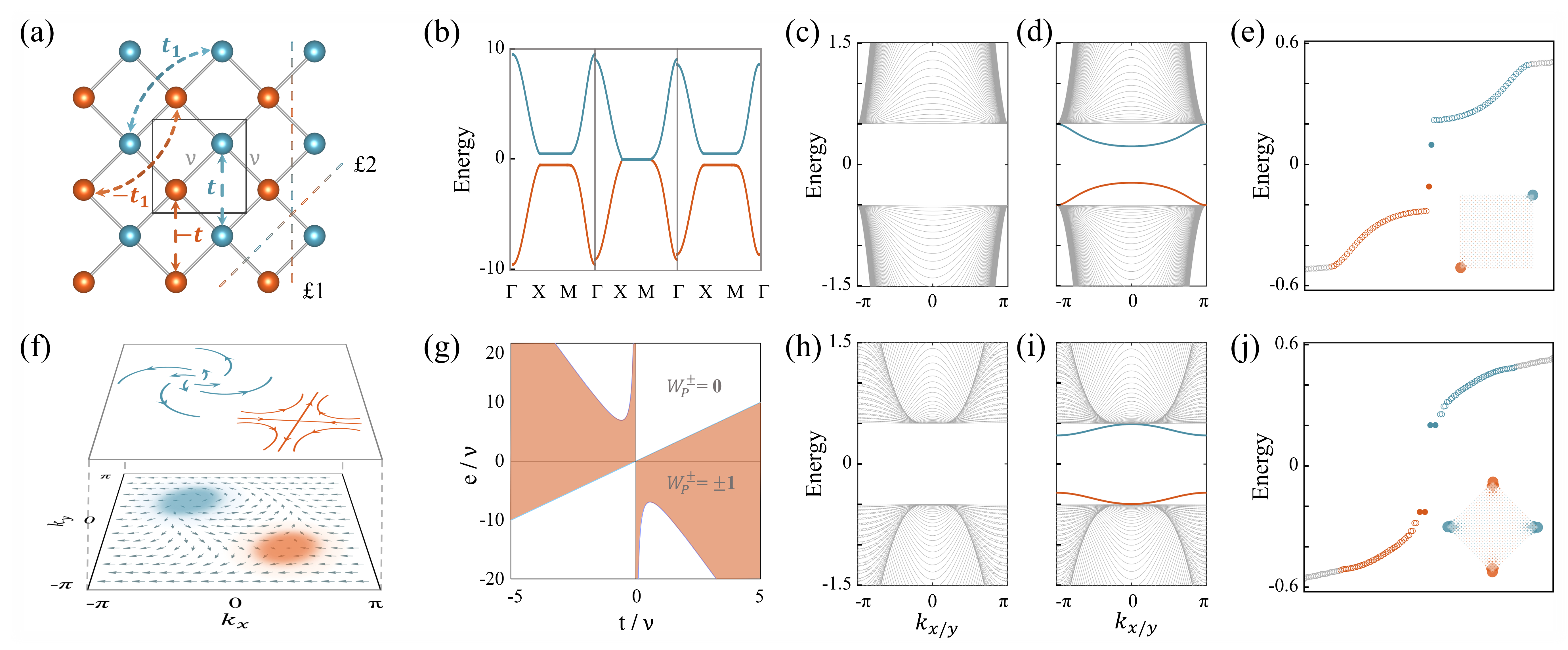}
	\end{center}
	\caption{(a) Schematic of the checkerboard lattice model. Not all hoppings are shown for clarity. (b) Band structure before and after the closure of the energy gap at $e=2$, with $\nu=t=1$, and $t_{1}=0.5$. (c-d) and (h-i) Edge spectra before and after the energy gap closure for type-I boundary $\pounds1$ and type-II boundary $\pounds2$ in (a). (e) and (j) The discrete energy spectra of the clusters corresponding to (d) and (i), respectively. The cluster size is $20\times20$ in the calculations. Here, $e=2.5$ for (c) and (h); $e=1.5$ for (d-e) and (i-j). (f) Vector $\mathbf{g}(k)$ in the Brillouin zone shown as a vector field. In the topological phase, $\mathbf{g}(k)$ exhibits a pair of topological vortex-antivortex structures. (g) A slice of topological phase diagram of the checkerboard lattice model at $ t_1 = \frac{1}{2}t$. The projected winding numbers $W_{2D,P}^{\pm} = \pm 1$ represent a non-trivial SOTI phase. }%
	\label{fig-3}%
\end{figure*}
As shown in the diagram in Fig. \ref{fig-1}(c), we can also understand it as projecting the origin $O$ into the plane where the closed curve formed by the trajectory of the endpoint of $\boldsymbol{d}(k)$ vector is located, obtaining the projected point $O_{g}$, and then calculating the winding number  of the $\boldsymbol{d}(k)$ around the projected origin $O_{g}$.
When the projection origin $O_{g}$ lies outside the closed curve, the system is considered to be trivial, whereas when it lies inside, the system is non-trivial (See Supplementary Materials (SMs) for more details). 

This model has two intriguing properties. First, its band gap may remain open during the topological phase transition as shown in Fig. \ref{fig-2}(a) with the spectrum evolving from $t<0$ to $t>0$. This feature is different from the typical topological phase transition, which involves a band gap closing and reopening. Note that there is no end state in the trivial regime with $0<t<0.5$. Second, during the generalized topological regimes in Fig. \ref{fig-2}(a) and (c), a pair of localized end states at the two ends of the chain are split with an energy difference,
  \begin{align}
 	\Delta E=2\frac{e\nu-2t\nu}{\sqrt{4 t^2+\nu^2}}.
 \end{align}
This splitting is due to the breaking of chiral symmetry, which differ from previous models that preserve chiral symmetry. The energy splitting of the two end states also exists in the Rice-Mele model \cite{RM}. The difference is that in the Rice-Mele model, staggered on-site energy only affects the amplitude of the end states splitting. In contrast, in our model, the staggered on-site energy plays a crucial role, not only influencing the amplitude of the splitting but also determining the topological nature of the model. Fig. \ref{fig-2}(c) provides the evolution of the energy spectrum as the staggered on-site energy changes. Fig. \ref{fig-2}(d) gives the phase diagram. The projected winding number $W_{1D,P}=1$ represents a non-trivial topological phase, with topological phase transitions determined by  
$e=2t$, $e=-2t-\nu^2/t$ and $t=0$. More details about the comparison of our model with the SSH and Rice-Mele models are presented in the SMs. 

\textit{2D model and topology}.-Now, we extend the model to 2D, and show that a SOTI phase could emerge. Taking a checkerboard square lattices as an example, we can also introduce a next-next-nearest-neighbor intra-sublattice hopping ($t_{1}$), as shown in Fig. \ref{fig-3}(a). The black labels represents the unit cell. In momentum space, the Hamiltonian of the two-band model of this checkerboard lattice can be written as,
\begin{align}
	H_{2D}(k)
	=\begin{pmatrix}
		h_{11} & h_{12} \\
		h_{12}^{\ast} & -h_{11}
	\end{pmatrix}.
\end{align}
Here, $h_{11}=e+2t(\cos(k_{x})+\cos(k_{y}))+4t_{1}\cos(k_{x})\cos(k_{y})$ and $h_{12}=\nu(1+e^{-ik_{x}}+e^{-ik_{y}}+e^{-i(k_{x}+k_{y})})$. Similar to the 1D case, the system undergoes two topological phase transitions with the evolution of the staggered on-site energy $e$, one of which has feature of closure of the energy gap. Upon consideration of  
$\nu=t=1$, and $t_{1}=0.5$,  it is observed that the closure of the energy gap occurs at $e=2$. Fig. \ref{fig-3}(b) shows the band structure before and after the energy gap closure. To determine the SOTI phase in 2D, the most intuitive method is to verify the existence of 1D edge states and 0D corner states. For a more comprehensive analysis, we investigate two different types of boundaries (Fig. \ref{fig-3}(a)), type-I boundary $\pounds1$ (constructed by the same sublattices) and type-II boundary $\pounds2$ (constructed by alternating sublattices), and construct square clusters containing integer unit cells on these boundaries.
Taking the type-I boundary $\pounds1$ as an example, two edge states can be observed in the bulk energy gap of the boundary spectrum when $e<2$ (Fig. \ref{fig-3}(d)). Accordingly, corner states can be observed in the clusters (Fig. \ref{fig-3}(e)). In this scenario, the system exhibits the SOTI phase. However, when $e>2$, only bulk energy bands are observed in the boundary spectrum and no corner modes emerge in the corresponding clusters (Fig. \ref{fig-3}(c) and SMs), indicating a trivial phase. The results are the same for the case with type-II boundary and cluster as well (Fig. \ref{fig-3}(h-j)). Thus, in the SOTI phase of this model, corner states are guaranteed to appear at corners formed independently of the boundary configuration. This distinguishes it from other models that rely on symmetries to exhibit corners states only at particular corners.

To characterize its generalized topological properties in 2D, we return to the parameter space. Similar to the 1D case, this model can also be written in the form of $H_{2D}(k) =\boldsymbol{d}(k)\cdot\boldsymbol{\sigma}$ with three components, After the projection operation, we can get a new 2D vector $\mathbf{g}(k)$. The projection operator $\hat{P}_{2D}$ is defined as follows,
\begin{align}
	\hat{P}_{2D} =\begin{pmatrix}
		\frac{\nu}{\sqrt{(t+2t_{1})^2+\nu^2}} & 0 & \frac{t+2t_{1}}{\sqrt{(t+2t_{1})^2+\nu^2}} \\
        0 & 1 & 0 \\
	    \end{pmatrix}.
\end{align}
Since the Brillouin zone is 2D, the trajectory of the vector $\mathbf{g}(k)$ forms a curved surface. We illustrate the $\mathbf{g}(k)$ in the Brillouin zone as a vector field, as shown in Fig. \ref{fig-3}(f). In the topological phase, the vector $\mathbf{g}(k)$ exhibits a pair of vortex-antivortex-pair structures centered at $\pm(k_{x_{0}}, k_{y_{0}})$. The topological structure of a vortex-antivortex pair can also be described by their phase winding numbers,

\begin{align}
      W_{2D,P}^{\pm}=
      \oint_{\mathcal{C}_{\pm}} \frac{d \vec{k}}{2 \pi}       \cdot\left[\frac{g_{x}}{|\vec{g}|} \vec{\nabla}\left(\frac{g_{y}}{|\vec{g}|}\right)-\frac{g_{y}}{|\vec{g}|} \vec{\nabla}\left(\frac{g_{x}}{|\vec{g}|}\right)\right] .
\end{align}
Here, "$\pm$" represent the upper and lower parts of the Brillouin zone, corresponding to the integration of the vortex and antivortex respectively. The result of winding number calculation in the topological phase is $W_{2D,P}^{\pm}=\pm 1$. This is also commonly referred to as a topological charge.
The transition of the system from trivial to non-trivial and back to trivial corresponds to the generation and annihilation of a topological vortex-antivorte pair. The evolution path of the center of vortices goes from $(0,0)$ to $(\pi,\pi)$ (see SMs for details). We plot a slice of the topological phase diagram of the checkerboard lattice model at $ t_1 = \frac{1}{2}t$ in Fig. \ref{fig-3}(g), where the characteristics of non-trivial and trivial topological phases are represented by the projected winding numbers $W_{2D,P}^{\pm} = \pm 1$ and $0$, respectively.

\begin{figure}[pt]
\begin{center}
\includegraphics[width=1.0\columnwidth]{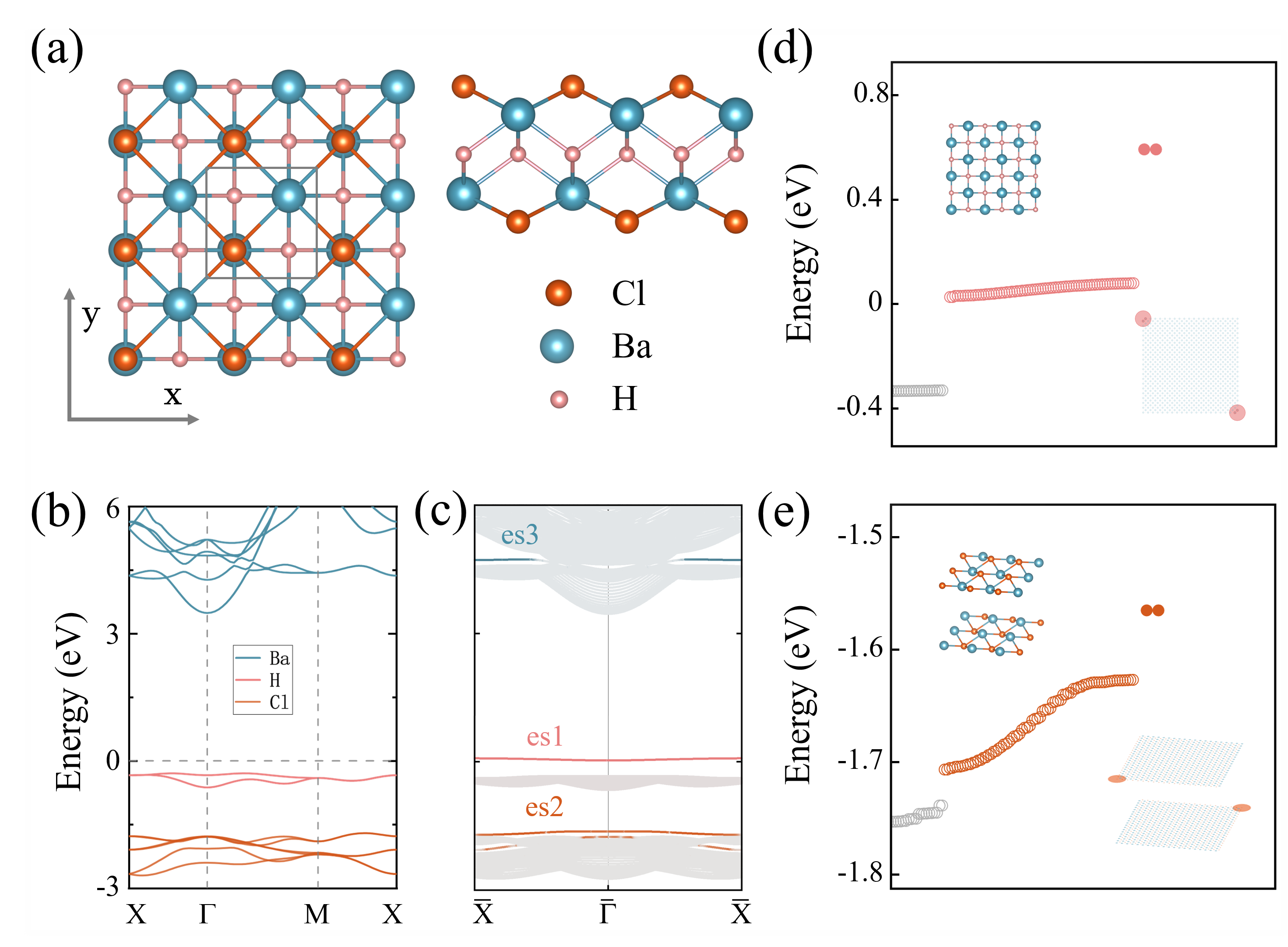}
\end{center}
\caption{(a) Top and side views of BaHCl. The black square represents the unit cell. (b) Bulk band structure of BaHCl, with different colors indicating the weights of different atoms . (c) Full edge spectra showing the three edge states $es1$, $es2$ and $es3$. (d) and (e) The discrete energy spectrum of the cluster. At different energy regimes, the corner states resulting from two sets of second-order topologies can be observed accordingly. (d) corresponds to nontrivial topology hosted in Ba-H-Ba layers. (e)  corresponds to nontrivial topology hosted in both top and bottom Ba-Cl layers. The cluster size is 20 × 20 in the calculations.}%
\label{fig-4}%
\end{figure}

\textit{Material realization}.-Now, we demonstrate that our model can be realized in many real materials (see SMs for details) and here we take BaHCl as an example \cite{https://doi.org/10.1002/zaac.19562830108,UBUKATA2022123253}.
BaHCl is a ternary hydride-halide adopting the matlockite structure with space group  $P4/nmm$. Fig. \ref{fig-4}(a) and (b) show the lattice and bulk band structure of BaHCl. Through first-principles calculations, we discovered that its five-layer structure (Cl-Ba-H-Ba-Cl layer) supports two distinct sets of second-order topology simultaneously, corresponding to different boundary configurations in our model.
The first set of SOTI arises between Ba and H, corresponding to the  middle Ba-H-Ba layer. It distinctly captures edge states $es1$ (Fig. \ref{fig-4}(c)) and corner states (Fig. \ref{fig-4}(d)) corresponding to boundary $\pounds2$ in our model.
The second set of SOTI originates between Ba and Cl, involving the inversion-symmetric top Cl-Ba layer and bottom Ba-Cl layer. It distinctly capturing edge states $es2$ (Fig. \ref{fig-4}(c)) and corner states (Fig. \ref{fig-4}(e)) corresponding to boundary $\pounds1$ in our model.
Despite the fact that the edge and corner states from Ba are embedded within the bulk states, we can effectively identify the SOTI by observing the edge and corner states of H and Cl. This approach mitigates challenges in future experimental observations. Finally, by determining the topological origin of the system based on the position of the last occupied state, we can distinguish it from other SOTI models.

\textit{Discussion and conclusion}.-We have proposed a highly versatile sublattice model without chiral symmetry, revealing its implicit topological information through projected winding numbers. Unlike most previous studies, this model no longer focuses on discussing the relative amplitudes of intracell hopping and intercell hopping, but instead treats them uniformly as nearest-neighbor hopping, which is common scenario in electronic materials. Consequently, this eliminates the dependence of the existence of end states and corner states, determined by bulk topology, on the selection of unit cells and boundary configurations, greatly reducing the experimental difficulty in sample preparation.

The presence of distinct end states and corner states localized at different sublattice terminations implies a significantly reduced requirement for the band structure of the material. By observing these states, we can discern the generalized topology between the  occupied/unoccupied band near the Fermi level and any unoccupied/occupied band.
In particular, the binary nature of end states and corner states here implies that the degenerate end states and corner states are necessarily fully occupied or unoccupied. This feature allows us to discern the topological origin and distinguish it from other models.  It is noteworthy that the sublattice in this model is not limited to different atoms but can also include orbitals, magnetism, etc., providing more possibilities for experimental observation and quantum device design.

\begin{acknowledgments}
We thank Chang-An Li and Song-Bo Zhang for helpful discussions. This work was financially
supported by the National Key R\&D Program of China (Grants No.
2022YFA1403200), National Natural Science Foundation of
China (Grants No. 92265104, No. 12022413, No. 11674331), the Basic Research Program of the Chinese
Academy of Sciences Based on Major Scientific Infrastructures (Grant No. JZHKYPT-2021-08), the CASHIPS Director’s Fund (Grant No. BJPY2023A09), the \textquotedblleft
Strategic Priority Research Program (B)\textquotedblright\ of the Chinese
Academy of Sciences (Grant No. XDB33030100), Anhui Provincial Major S\&T Project (Grant No. s202305a12020005), and  the Major Basic Program of Natural
Science Foundation of Shandong Province (Grant No. ZR2021ZD01). A portion of
this work was supported by the High Magnetic Field Laboratory of Anhui
Province, China
\end{acknowledgments}

\end{document}